\documentclass{elsart}

\usepackage{harvard}

\usepackage{graphicx}

\usepackage{amssymb}

\def\url#1{{\ttfamily\def\/{/\discretionary{}{}{}}#1}}

\begin{document}

\begin{frontmatter}
\title{The Problems of Ageing}


\author[Rudnick]{Lawrence Rudnick\thanksref{lr}}

\thanks[lr]{E-mail:  larry@tc.umn.edu}

\address[Rudnick]{Department of Astronomy, University of Minnesota, 116 
Church St. SE, Minneapolis, MN 55455-0149}

\begin{abstract}
I present a brief overview of observational, modeling and
theoretical issues related to ageing calculations based on 
spectral steepening. These include problems such as inhomogeneous
magnetic fields, diffusion of relativistic particles, confusion
from multiple particle populations and particle acceleration.
Although some of the effects are only of order unity, others
call into question the entire ageing paradigm.  I refer to and show some
data illustrating these problems and make a few recommendations
about how we should proceed given these uncertainties.
\end{abstract}
\end{frontmatter}

\section{Introduction}
\label{intro}
The relative depletion of high energy relativistic electrons due to
radiative losses, either by the synchrotron or Inverse Compton 
mechanisms, is a well-understood physical process.   It is likely 
to explain the almost universal observation that at sufficiently
high frequencies, the broadband spectra of extragalactic radio 
sources fall below the power-law extrapolation from lower 
frequencies.  From this observation, with a large number of 
assumptions, it is possible to derive 
an age for the relativistic electron population, usually 
representing the time since the particles were last accelerated.	

Before addressing the problems with how such age estimates are made, 
let's first look at why, and at what level of accuracy, age estimates
become important.  This issue has received little, if any, discussion 
in the literature.  Most importantly,  radiative ages
provide a basic confirmation and constraint on the physical models
of extragalactic sources.  E.g., where particle lifetimes are demonstrably
short, we typically invoke {\it in situ}  particle acceleration.  This leads
us to physical models including shocks or turbulence or ``freezers" 
\cite{freeze}.    
Radiative ages, whether short or not,  can also be useful for 
comparison with estimates derived from other methods, e.g., 
energy supply, dynamics, or evolutionary models.  For all of the 
above, radiative ages within a factor of a few are probably sufficient.

\section{Overview of Ageing Problems}
\label{problems}
To understand the fundamental ways in which ageing calculations may be 
wrong, it is useful to characterize the standard and alternative pictures of the 
evolution of the relativistic electron population.  
In the {\bf standard} model, particles are accelerated to high energies in nucleus, 
jets, and hot spots.  {\it Downstream steepening of the spectra indicate radiative 
energy losses} of an initially homogeneous electron population.
In an {\bf alternative} picture, particles that are injected to the downstream flow from 
nucleus, jets and hot spots have a range of particle acceleration histories;  all 
have already experienced radiative losses.  {\it Downstream steepening or 
flattening occurs through magnetic field and adiabatic variations, further particle 
acceleration, inhomogeneities and perhaps further radiative losses} (although 
the last may not be necessary).

Observationally, there is a fundamental degeneracy in that many 
plausible factors can 
mimic the appearance of radiative losses.  
In such cases, standard ageing analyses can be made but would have no 
meaning.    
Steepening of the radio spectrum with distance from the nucleus or a hot spot 
 might be due to overall variations in the magnetic field strength, 
adiabatic losses of electron energies and field strengths, changes in the filling 
factor of high field regions, the onset or decline of efficient particle acceleration, 
changing mixtures within a beam of otherwise non-evolving particle 
populations, in addition to the almost universally invoked radiative losses. 
There is increasing evidence, on both theoretical and observational grounds, 
that these confusing effects are not simply pessimistic speculations but actually 
occur. 

It is important to remember that most measurements of ageing are based on a 
{\bf second order} effect, namely the change in the observed spectral index from one 
location in the source to another.  So while it is likely that losses at some time 
and place have probably caused the observed curvature in the spectrum, it is 
much less clear whether local variations in spectral index reflect subsequent 
particle energy losses and gains.

Most of the ideas presented here are not new.  Questions regarding the validity of 
radiative age estimates have a long history of being proposed and forgotten, 
largely because of the complications of dealing with them. 
Quick on the heels of the discovery of M87's optical synchrotron jet, 
\citeasnoun{fel68}
pointed out that relativistic particle acceleration must take place 
{\it in situ} (cf \cite {heinz} ). 
By the mid-70s, the increasing number of bridges with spectral gradients
allowed ageing analyses to become popular (e.g., \citeasnoun{jenk76}).  During this period \citeasnoun{wil78} suggested
the need for {\it in situ} acceleration in giant radio galaxies.

Enormous progress was made in spectral mapping programs in the 1980s.  
Among the problems raised, \citeasnoun{alex87} pointed out that 
derived ages did not extrapolate to zero at the hotspots, as they should.  They 
suggested that adiabatic expansion 
losses would shift the break frequency 
and mimic a finite age immediately post-hot spot. It is disappointing to note that no 
one has yet taken up the necessary calculation of the radiative evolution of a 
curved injection spectrum.

The first serious treatments of how temporal changes in the magnetic field or 
inhomogeneities in the field can affect spectral ages were carried out 
by  \citeasnoun{wit90}  and \citeasnoun{siah90}. \citeasnoun{blun94} introduced 
a method for interpeting spectral breaks when both adiabatic and
radiative losses are present. Most of this work has been virtually 
ignored in the literature.
During the 1990s, a few workers have explored the relationship between source 
structure, dynamical ages, and radiative ages. \citeasnoun{eil96} has argued
that radiative ages are deceptively short, such as will result from
emission in highly inhomogeneous magnetic fields \cite{eil97}. Tribble (1993,
1999) has investigated the effects of random magnetic fields on ageing
calculations and the effects of electron diffusion on the spectral shape.
\citeasnoun{kai97} have looked in detail at the energy {\it history} of radiating
particles, another major oversight in the standard analyses.

\section{Evidence for Confusion in Radiative Age Calculations}
{\bf The magnetic field and apparent ageing.}
We start with the oft-ignored fact that at a fixed observing 
frequency, $\nu_{obs}$, we 
sample particles with a characteristic energy $\gamma_{obs}$  depending on the local value 
of the magnetic field, $B_{local}$, according to $\nu_{obs}  = 
\gamma^{2}_{obs} B_{local}$ .  When the local field 
changes, so does the energy of the observed particles and their slope 
(the spectral index) if the underlying electron population is already 
curved.  This generally implies without any additional losses, a consequence of
lower $B_{local} $ will be both a decrease in emissivity {\it and} a 
steepening of the spectral index.  This 
correlation of low brightness and steep spectral index is almost universal, and
is strong even at low frequencies where the emissivity is expected to 
be insensitive to losses (e.g., Leahy, Muxlow and Stephens, 1987).  
Variations in $B_{local}$ thus stand as a viable alternative to local ageing.  What is 
needed is a serious, quantitative estimate of the possible effects of these 
variations in each source where a claim of radiative ageing is to be made.  
When we attempted to do this for the steep-spectrum sheaths in two WATs, we 
were unable to distinguish between variations in $B_{local}$ and different radiative 
loss histories \cite{kat99}.

Perhaps the most dramatic example of this problem is seen in the beautiful work 
of \citeasnoun{fer98} on the large scale structure of NGC 1265, who show a
show a plot of intensity and spectral index as a function of distance from the 
core of this head-tail source.  At a distance of 13 arcmin, the intensity drops 
precipitously by almost an order of magnitude and the spectral index suddenly 
steepens from $-1.2$ to $-2.2$.  These changes are certainly unrelated to ageing, 
and likely result from a sudden drop in the magnetic field strength along the 
flow or a completely separate electron population.  When one looks at the more 
gradual brightness declines and spectral steepening in the first 13 arcmin, it is 
then proper to ask whether gradual radiative losses or magnetic field reductions 
or both are responsible.

Another major problem arises from our untested assumption that the 
energy densities of relativistic electrons track those in the magnetic 
fields. This directly influences our calculations of magnetic field 
variations and radiative ages across a source.
There are a variety of circumstances in which this would not 
be true. We are on the cusp of actually measuring the magnetic field 
distributions in extragalactic sources through the study of Inverse Compton 
emission as suggested by  \citeasnoun {har78}. Early results on Centaurus B by \citeasnoun{tash98}
and on 3C219 by
\citeasnoun{bru99} suggest that variations in the relativistic electron
density (relative to the magnetic field energy) may need to vary by factors
of $2-3$ to explain the radio/X-ray ratios.  Although these conclusions
are not robust at present, they do suggest a possible major flaw in our
assumptions of how magnetic fields vary within a source.

{\bf Misleading Clues to Particle Acceleration.}
I was shocked, as all observers should be, when I first saw the 
results of the numerical simulations described in these proceedings 
by \citeasnoun{tre99}.  They found
that there was little correlation between regions of 
flat spectra and regions of high emissivity. Since this is not what
we expect, e.g., because we expect shocks to be bright and to accelerate
particles to flatter spectra, and not what we observe, which is the
almost universal correlation between bright regions and flat spectra, 
their result deserves some explanation.

The regions of flat spectra in the simulations are those where 
electrons have passed through shocks.  In other words, the spectrum
of the particles in a given location is a reflection of the {\bf 
history} of those particles, {\bf not} simply the local hydrodynamic
conditions at the time of observation.  Therefore, it is possible 
to have regions of flat spectrum with no cospatial shock.  Similarly,
they find that shocks can be currently small and insignificant in emissivity
but still have produced a distinct downstream trail of flat spectrum particles 
which may be observed.   We must therefore approach the identification
of flat spectrum regions and particle accleration sites with great
caution. This also implies that particle acceleration can confuse
measurements of radiative loss rates even when no site for the 
acceleration is visible.

We have the remaining question of why bright regions are always
observed to have flatter spectra if this is not seen in the simulations.
In the simulations \citeasnoun{tre99} have performed
so far, radiative losses are not important.  
Therefore, the spectra are straight at each location, and changes in
magnetic field along will not change the observed spectral slope.  As
discussed above, however, when the electron distribution is already
curved, high magnetic fields will produce both high emissivities and
flat spectra, independent of any particle acceleration.

{\bf Inhomogeneities.} Inhomogeneities in the magnetic fields can have important consequences
both for the evolution of the relativistic particles and for how they
appear in observations.   On the physical side, \cite{eil97} and 
\cite{tripre} have discussed the evolution of electron populations 
when there is diffusion between regions of different magnetic field 
strengths.  The spectra can behave much differently than the standard 
homogeneous models, and ageing calculations in some circumstances can
be completely misleading.

On the observational side, we have shown through the use of spectral 
tomography (e.g., \citeasnoun{kat97},\citeasnoun{kat99},\citeasnoun{rud96})
that many sources have overlapping structures of different spectral
indices.  Figure 1 shows one example.  When these overlapping 
structures are blended and spectral variations are measured along a 
source, ageing calculations are meaningless.  Using tomography or 
another method prior to ageing analysis should be {\it de riguer}.

\begin{figure}
\begin{center}
\rotatebox{0}{\scalebox{.7}[.7]{\includegraphics[3cm,10cm][20cm,23cm]{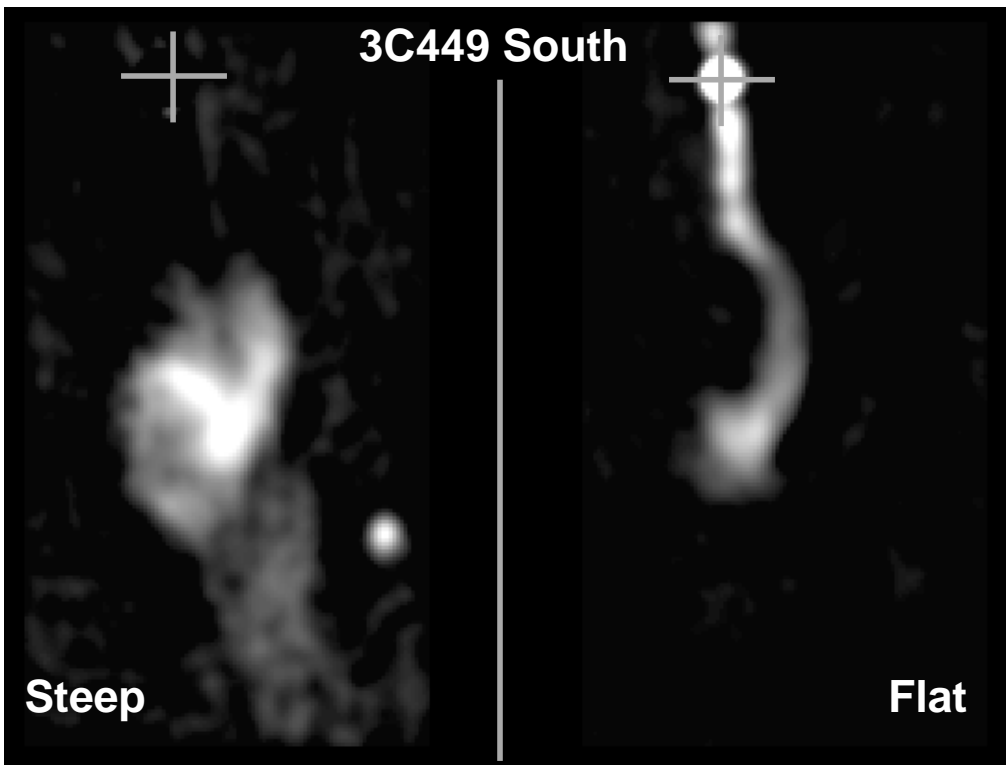}}}
\end{center}
\caption{
Two frames from the spectral tomography gallery between $\lambda 6cm$ and
$\lambda 20cm$ for 3C449 (Katz-Stone \& Rudnick, 1997).  On the left(right) we have isolated the
steeper(flatter) spectrum material. Where the flat jet is often mistakenly
thought to flare, we see that a new steep spectrum component appears. 
Spectra that blend these two structures are not appropriate for ageing
calculations.}
\label{tomo449}
\end{figure}

When there is insufficient resolution to detect blends, or the blending
of components occurs primarily along the line of sight, the shape of 
the spectra can still be used to assess its importance.  Despite 
numerous claims to the contrary in the literature, we have yet to
find a single source where the spectra are well-fit by any of the
standard ageing models.  Although data at a given position may appear
consistent with one model, nearby positions would yield an apparently
different low frequency index, or a different shape, etc.  All of this
can be assessed using color-color diagrams, as introduced by 
\citeasnoun{kat93}.  Figure 2 shows how color-color diagrams are constructed
and used, and how the two tails of 3C465, e.g., have quite different 
shapes. These differences are probably a result of either superposition or diffusion
effects.  \citeasnoun{young99} show results on other WATs.
In such cases, ageing calculations can only be performed 
after the confusing effects have been removed;  this is currently
beyond our capability in most cases.

\begin{center}
\begin{figure}
\rotatebox{0}{\scalebox{.7}[.7]{\includegraphics{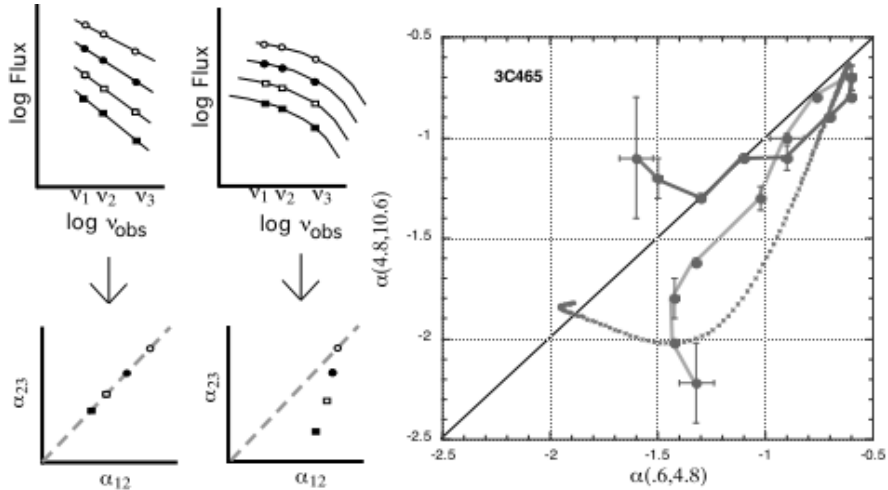}}}
\caption{Color-color diagrams.  On the left are two illustrations of 
how spectra in normal [log I, log $\nu$] space translate into color-color
space. Left cartoon: different power laws.  Right cartoon: the same curved spectrum 
at each location, but {\it shifted} in frequency as would
appear under magnetic field changes or ageing.
On the right is the color-color diagram along the two tails of
3C465 from the data of Feretti \etal, (1998).  The best standard spectral
shape, a KP spectrum, is shown as a broken line.  No standard shapes fit and
the meaning of any ageing analyses is unclear.}
\label{cc465}
\end{figure}
\end{center}

\section{Closing Remarks and Recommendations}
At this meeting, there was extensive discussion about how pernicious selection 
effects can be in radio source population studies.   Similar problems can lead to 
spurious correlations between radiative age estimates and other derived source 
properties such as jet power or dynamical ages. {\it Caveat Emptor}.

What should we do in the face of all these serious questions about the
validity of radiative ages?  I feel that two approaches are important.

\begin{itemize}
	
\item{{\it Don't ignore the discrepancies}-- when the spectra do not 
steepen monotonically, when the low frequency index appears to change
throughout a source, when the standard spectral shapes don't fit.} 
\item{{\it Evaluate alternative histories carefully.}  This means 
that in each source we need to specifically ask what evidence or limits can be
placed on a) spectral variations due to magnetic field and/or 
adiabatic changes, b) {\it in situ} particle acceleration, and c) 
confusion between overlapping particle populations.}
\end{itemize}

In summary, it is not sufficient to perform a standard radiative loss analysis and
assume that you have learned anything about the age of a source.

This work is supported by NSF grant AST 96-16964.  I am grateful for 
many stimulating conversations with T. Delaney, J. Eilek, M. Hardcastle, T. Jones, D. Katz-Stone, 
B. Koralesky, J.P. Leahy, I. Tregillis, A. Young and others who wish to protect 
their reputations by remaining anonymous.

{\bf Q:} (Peter Barthel)  Which age esimates do you trust?  Don't we agree that 
size/0.1c and spectral ages are in the same ballpark?\\
{\bf A:} Yes, but the agreement between these two numbers is not meaningful to me. 
There are many subtle selection effects that can give spurious 
correlations.  Measured advance speeds, as discussed at this meeting by Conway, 
probably give reliable order of magnitude ages for at least the CSS 
sources.\\ \\
{\bf Q:} (Dan Harris)  I don't think you should use current claims of IC/3K X-ray 
emission as evidence for relativistic electron populations in extremely low fields.  
Therefore, I question your claim of ``field not tracking particles."\\
{\bf A:} I agree that the current observations are not conclusive.  
However, the most straightforward interpretation of the data is that the ratio of 
relativistic particle/field energies is not constant.  This says we need to take this 
issue seriously.

\end{document}